\documentclass[aps,prfluids,reprint,unsortedaddress,onecolumn,longbibliography]{revtex4-2} %preprintnumbers, superscriptaddress
\usepackage{hyperref}
\usepackage{amsmath,latexsym}
\usepackage{graphicx}
\usepackage{siunitx}
\usepackage{textcomp}
\usepackage{wasysym}
\usepackage{amssymb}
\usepackage{MnSymbol}
\usepackage{oplotsymbl}
\usepackage{physics}
\usepackage{array}
\usepackage{tikz}

\DeclareMathOperator\erfc{erfc}

\begin{document}
\title{Lasting effects of discontinuous shear thickening in cornstarch suspensions upon flow cessation}
\date{\today}
\author{Jae Hyung Cho}
\thanks{J.H.C. and A.H.G. contributed equally.}
\email[Corresponding author. ]{jaehcho@mit.edu}
\author{Andrew H. Griese}
\thanks{J.H.C. and A.H.G. contributed equally.}
\affiliation{Department of Mechanical Engineering, Massachusetts Institute of Technology, Cambridge, Massachusetts 02139, USA}
\author{Ivo R. Peters}
\affiliation{University of Southampton, Faculty of Engineering and Physical Sciences, Highfield, Southampton SO17 1BJ, UK}
\author{Irmgard Bischofberger}
\email[Corresponding author. ]{irmgard@mit.edu}
\affiliation{Department of Mechanical Engineering, Massachusetts Institute of Technology, Cambridge, Massachusetts 02139, USA}

\begin{abstract}
Dense suspensions that exhibit discontinuous shear thickening (DST) undergo complex stress relaxation when the flow abruptly stops. Using rotational rheometry, we study the two-step relaxation of aqueous cornstarch suspensions out of the DST state upon flow cessation and show that the DST fluid retains the memory of its shear-thickening state until the shear stress reaches a constant value at late times. We find that this residual stress at the end of the relaxation increases with the steady-state viscosity before the cessation. Furthermore, the timescales that characterize the two-step exponential decay of the shear stress exhibit near linear dependence on the steady-state viscosity. Within the current framework that ascribes DST to the breakdown of hydrodynamic lubrication layers leading to interparticle frictional contacts, the lasting effects of the steady-state viscosity suggest that the memory of frictional contacts persists until the end of the relaxation, despite the presence of repulsive forces between the particles. These results indicate that complete, spontaneous relaxation of the system out of DST is stalled by the partial retention of the frictional force chains, which may be caused by the stationary boundaries and the adhesion between cornstarch particles.
\end{abstract}

\maketitle

% Introduction
\section{Introduction}
When sheared at controlled rates, a concentrated suspension of Brownian or non-Brownian particles can exhibit a sharp increase in the apparent viscosity, defined as the macroscopic stress divided by the global shear rate, by orders of magnitude at a critical shear rate. Such discontinuous shear thickening (DST) is attributed to the transition from hydrodynamic to frictional interactions between the particles at sufficiently high packing fractions, which proliferates as the stress applied exceeds that due to inherent interparticle repulsions \cite{Lootens2005,Brown2012,Seto2013,Mari2014,Guy2015a,Lin2015,Royer2016,Clavaud2017,Comtet2017,Morris2020}. At applied stresses lower than the repulsive stress, adjacent particles interact via lubrication forces. At applied stresses higher than the repulsive stress, the lubrication layers break down and the particles come into frictional contact. When the applied stress is sufficiently high such that the frictional forces dominate particle interactions, a system-spanning contact network resists shear by continuously breaking and reforming \cite{Seto2013,Mari2014}. Compared to lubricated interactions, frictional interactions between particles cause the jamming transition to occur at lower packing fractions \cite{Boyer2011}. The emergence of frictional interactions, therefore, effectively brings the system closer to the packing fraction at jamming, which induces the discontinuous increase in the apparent viscosity at a critical shear rate \cite{Wyart2014,Peters2016}. When sheared at controlled stresses, this lubricated-to-frictional transition manifests itself as the apparent viscosity linearly increasing with the stress. Once the applied stress exceeds the onset stress at which the lubrication films start to rupture, frictional contacts between the particles become more prevalent with increasing applied stress, which leads to the gradual increase in the apparent viscosity \cite{Brown2012,Guy2015a,Royer2016,Peters2016}. \par

This description of DST, however, primarily concerns steady-state flows. Although macroscopic features of various transient flows of dense suspensions are successfully captured by continuum models \cite{Baumgarten2019,Gillissen2019}, the pathways to the formation of a contact network upon the onset of shear or the relaxation of the network upon the removal of shear remain nebulous at the particle level. These kinetic processes introduce characteristic timescales of the transitions into and out of DST, which can give rise to oscillatory behaviors at macroscopic levels in the presence of competing external timescales \cite{vonKann2011,Richards2019}. A sound understanding of the transient phenomena is thus essential for avoiding \cite{Toussaint2009} or exploiting \cite{Lee2003,Gurgen2017} DST in industrial applications. Extensive explorations of the solidlike behaviors of dense suspensions due to a dynamic jamming transition, such as impact resistance \cite{Waitukaitis2012,Jerome2016,Maharjan2018} and crack propagation \cite{Roche2013}, have revealed the roles of particle inertia, pressure of the suspending fluid, and boundary conditions. DST fluids upon flow cessation yield signatures of complex stress and microstructural relaxations, whose mechanisms are left largely unexplored \cite{DHaene1993,OBrien2000,Lootens2005,Maharjan2017,Baumgarten2020}. \par

Specifically, recent studies report the presence of at least two steps of stress relaxation, which hints at a nontrivial interplay between viscous, elastic, and plastic events on microscopic scales \cite{Maharjan2017,Baumgarten2020}. In their experimental study, Maharjan and Brown report multistep stress relaxation of cornstarch suspensions upon flow cessation, where the stress initially drops rapidly, and then exhibits a slower two-step exponential decay to a nonzero plateau \cite{Maharjan2017}. Baumgarten and Kamrin employ a constitutive model \cite{Baumgarten2019} that captures the behavior reported in Ref. \cite{Maharjan2017} to unveil underlying viscoelastic and viscoplastic responses \cite{Baumgarten2020}. Although the interpretations of their results vary, both studies suggest that the sequence of relaxation steps is associated with the long-lasting system-spanning contact network. \par

In this work, we use rotational rheometry to investigate the relaxation of dense aqueous cornstarch suspensions out of DST upon flow cessation and demonstrate that the effect of DST lasts until the end of the relaxation. We show that the steady-state viscosity $\eta_0$ in DST is positively correlated to the residual stress $\sigma_\infty$ at the end of the relaxation and that $\eta_0$ determines all the timescales that characterize the relaxation. These findings reinforce the notion of the lasting effects of the contact network in flow cessation. Unlike in the rheological study by Maharjan and Brown that probes the relaxation of the dense suspensions initially sheared at a rate above the critical shear rate \cite{Maharjan2017}, we shear the sample at a constant stress $\sigma_0$ to ensure that the system remains at or very close to the critical shear rate $\dot{\gamma}_c$, at which the apparent viscosity $\eta_0$ discontinuously increases, during the steady-state flow. Because the apparent viscosity $\eta_0$ does not saturate to a maximum value beyond the DST regime in our experiments, our protocol of applying a controlled stress allows us to pinpoint the state of the samples during shear to be exactly within the DST regime, where the lubricated-to-frictional transition is underway. Although the sample could in principle be directly sheared at the critical shear rate $\dot{\gamma}_c$, the determination of its exact value is often unfeasible because of its sensitivity to sample age and shear history, which we cannot perfectly control. Even if imposing the exact critical shear rate $\dot{\gamma}_c$ were possible, the resultant apparent viscosity would suffer large fluctuations due to the very discontinuity \cite{Lootens2003,Sedes2020,Rathee2020}, which calls for a stress-controlled procedure for the system to reach a steady state. \par
  
Once the steady state is reached, the shear rate $\dot{\gamma}$ is abruptly set to zero, upon which the shear stress $\sigma(t)$ exponentially decays with time $t$ in two distinct steps to a late-time nonzero constant $\sigma_\infty$. Remarkably, we find that the residual stress $\sigma_\infty$ generally increases with the apparent viscosity $\eta_0$ measured during the steady-state flow prior to the flow cessation, although no precise relation between $\sigma_\infty$ and $\eta_0$ can be ascertained owing to the significant amount of scatter in our data. In addition, we observe that both characteristic times of the two-step exponential decay linearly increase with the apparent viscosity $\eta_0$, as long as the particle mass fraction and the applied stress $\sigma_0$ are sufficiently high such that the system is initially sheared well into the DST regime. The transition time between the two steps and the time at which the stress $\sigma$ reaches its late-time constant $\sigma_\infty$ also increase nearly linearly with $\eta_0$, exhibiting power laws with exponents close to unity. This dependence of the residual stress $\sigma_\infty$ and the time parameters of the stress relaxation on the apparent viscosity $\eta_0$ suggests that the system retains its memory of the level of frictional contacts prior to the flow cessation even after the relaxation ends, despite the presence of the interparticle repulsions that enable DST. We infer that the frictional force chains in the system-spanning contact network formed during shear partially persist after the flow stops by maintaining a fraction of the frictional contacts. This incomplete relaxation of the force chains may result from the geometric frustration of the system due to the stationary boundaries imposed and the adhesion between the cornstarch particles in water, which comes into play only after non-lubricated contact is made \cite{Galvez2017}. \par

% Methods
\section{Methods}
\subsection{Sample preparation}
We prepare dense suspensions by manually mixing cornstarch particles (Product number S4126, Sigma-Aldrich) with water until the sample appears homogeneous and its surface glossy. The cornstarch particles have irregular shapes and are highly polydisperse; their mean diameter and density are approximately $13\;\si{\micro\meter}$ and $1.6\;\si{\gram\per\cubic\centi\meter}$ \cite{Madraki2017}. Despite the significant density difference between the particles and water, we confirm that our results are not affected by the sedimentation by repeating selected experiments with density-matched samples for which a $37.5\,\textrm{wt}\%$ cesium chloride (Bio-world) solution is used instead of water. We report the concentration of the samples in terms of the mass fraction of the particles since their porosity and the water content cannot be readily measured \cite{Brown2012}. In addition, we find that the steady-state apparent viscosity of the suspensions made of particles from different batches differs systematically. Thus, to use a rheologically consistent measure of the concentration, we first obtain the steady-state apparent viscosity as a function of the stress to identify the minimum apparent viscosity of samples at different nominal mass fractions for each batch. We then scale the nominal mass fraction by a constant, such that the minimum apparent viscosity versus the scaled mass fraction curve collapses onto a reference curve. We report the scaled mass fraction as the effective mass fraction $\phi_m$, which takes into account the batch-to-batch variation. We use samples of effective mass fractions $\phi_m$ in the range of $58.3\% - 60.0\%$, which corresponds to the concentrations that are deep within the DST regime but are sufficiently low such that the samples can be reliably handled. \par

\subsection{Rheometry}
We use a stress-controlled rheometer (DHR-3, TA Instruments) with a parallel plate geometry of diameter $2R=40\;\si{\mm}$, coated with sandpaper (grit size: P240, $58.5\;\si{\micro\meter}$ in diameter) to prevent wall slip. The high concentration of the suspensions requires that the samples are loaded with care by gradually closing the gap while constantly wiggling the rod to avoid premature jamming of the particles. We set the final gap size $h$ to be $1250\;\si{\micro\meter}$, and ensure that the samples are in relaxed states by wiggling the rod after reaching the final gap size. Some experiments are repeated with a smaller gap size $h=1000\;\si{\micro\meter}$, and we confirm that there is no systematic change in the results. To achieve a consistent initial condition, we preshear each sample at a shear rate of $0.08\;\si{\per\s}$ for $30\;\si{\s}$, and perform equilibration at zero shear rate for another $30\;\si{\s}$. The temperature is fixed at $22.7\si{\celsius}$. \par

\begin{figure}[b]
\setlength{\abovecaptionskip}{-20pt}
\hspace*{-0.06cm}\includegraphics[scale=0.12]{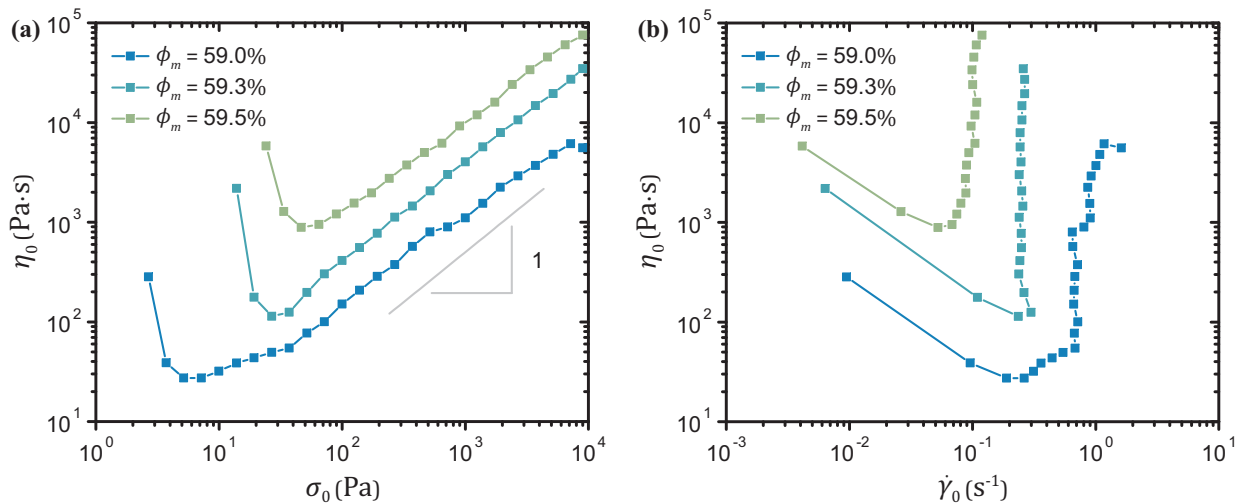}
\caption{\label{flowcurves} Steady-state apparent viscosity $\eta_0$ of dense cornstarch suspensions for three effective mass fractions $\phi_m$ as a function of (a) the applied stress $\sigma_0$ and (b) the shear rate $\dot{\gamma}_{0}$. Discontinuous shear thickening (DST) at a nearly constant shear rate is observed for each $\phi_m$.}
\end{figure}

To characterize the rate dependence of the steady-state apparent viscosity $\eta_0$ of our samples, we perform stress sweep experiments by gradually increasing the applied stress $\sigma_0$ using an equilibration time of $5\;\si{\s}$ and an averaging time of $4\;\si{\s}$. All samples exhibit shear thinning at low stresses $\sigma_0$, typical of dense cornstarch suspensions \cite{Brown2012,Peters2016,Maharjan2017,Madraki2017}, as shown in Fig.~\ref{flowcurves}(a,b). This shear thinning is followed by DST, where $\eta_0$ increases linearly with $\sigma_0$, as displayed in Fig.~\ref{flowcurves}(a), which is equivalent to an increase in $\eta_0$ at a nearly constant shear rate $\dot{\gamma}_0=\sigma_0/\eta_0$, as shown in Fig.~\ref{flowcurves}(b). Unlike some other studies of cornstarch suspensions that report another shear-thinning regime at high stresses beyond the DST regime \cite{Brown2012,Fall2015,Maharjan2017}, we do not observe a definite end of DST at high $\sigma_0$ in our samples. \par  

In the transient stress relaxation experiment, we initially apply a constant stress for $10\;\si{\s}$ in the range of $\sigma_0=1-11\;\si{\kilo\pascal}$ that lies in the DST regime for all $\phi_m$. Since our samples in steady state show DST at constant shear rates, we interpret $\dot{\gamma}$ measured at the end of this constant-stress flow step to be the critical shear rate $\dot{\gamma}_c$ and the corresponding apparent viscosity to be the steady-state apparent viscosity $\eta_0$. After this initial step, we abruptly set the shear rate $\dot{\gamma}$ to zero at the step time $t=0\;\si{\s}$ to observe the resultant relaxation in terms of the shear stress $\sigma_{{\theta}z}(t)\equiv\sigma(t){\equiv}2T(t)/\left({\pi}R^3\right)$, and the mean normal stress $\sigma_{zz}{\equiv}F_{z}(t)/\left({\pi}R^2\right)$, where $T$ and $F_z$ denote the torque and the axial force, respectively. The shear rate $\dot{\gamma}(t){\equiv}{\omega(t)}R/h$, where $\omega$ denotes the angular speed, is also monitored. A sample loaded in a plate-plate geometry experiences a spatially variant shear rate, and $\dot{\gamma}(t)$ defined above represents the maximum shear rate at the edge of the geometry. To account for the high stiffness of the samples in DST, we set the motor mode to stiff during this rate-controlled step. \par

\subsection{Relaxation characterization}

\begin{figure}[b]
\setlength{\abovecaptionskip}{10pt}
\hspace*{-0.06cm}\includegraphics[scale=0.12]{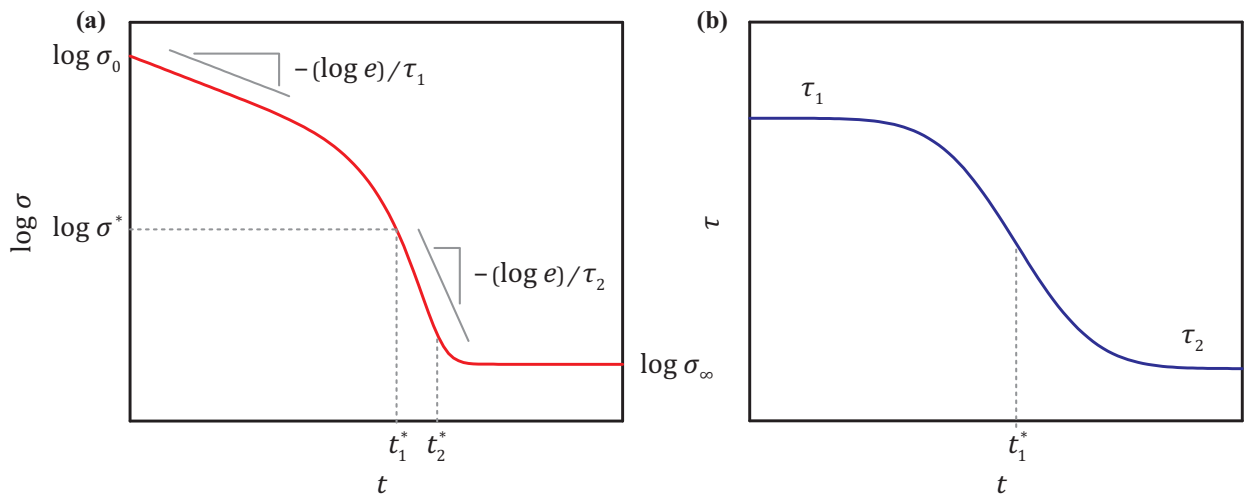}
\caption{\label{fit_schematic} (a) Schematic of our empirical model of the shear stress $\sigma(t)$ in the log $\sigma$-linear $t$ scale with relaxation parameters indicated. (b) Schematic of the complementary error function in the linear $\tau$-linear $t$ scale.}
\end{figure}

For most of our experiments, the shear stress $\sigma(t)$ upon flow cessation closely follows a stepwise exponential decay with two distinct characteristic times to a finite plateau, similar to what has been reported in Ref.~\cite{Maharjan2017}. To quantify this relaxation behavior, we use an empirical model that captures these features, as schematically shown in Fig.~\ref{fit_schematic}(a). We determine the parameters of the model by fitting an exponential function 
\begin{equation}
    \sigma(t)=\sigma_0\exp\left(-\frac{t}{\tau(t)}\right)+\sigma_{\infty}, \label{fit_function}
\end{equation}
which allows the characteristic time $\tau(t)$ to change from one value to another following a time-dependent complementary error function
\begin{equation}
    \tau(t) = \frac{\tau_1-\tau_2}{2}\erfc\left(\frac{t-t_1^*}{T}\right)+\tau_2 \label{fit_function_time}
\end{equation}
that changes from $\tau_1$ to $\tau_2$ at the transition time $t_1^*$, as shown in Fig.~\ref{fit_schematic}(b). Among various sigmoidal functions, the complementary error function is chosen as $\tau(t)$ smoothly varies from one plateau to another in the linear $t$-scale. The transition time between the two steps of exponential decay is denoted by $t_1^*$, and the scale parameter $T$ governs the rate at which the transition occurs. \par

We first identify the value of $\sigma_\infty$ from the mean of $\sigma(t)$ in the late-time plateau. We then perform a linear fit to the early time $\sigma(t)-\sigma_\infty$ in the log-linear scale and assume the $y$-intercept of the linear fit to be $\sigma_0$. Since the first exponential decay of $\sigma(t)-\sigma_\infty$ typically starts after a brief time lag, this adjusted $\sigma_0$ is slightly greater than the actual stress applied during the preceding constant-stress flow. The determination of $\sigma_\infty$ and $\sigma_0$ allows us to invert Eq.~\eqref{fit_function} to obtain $\tau(t)$ to which we fit Eq.~\eqref{fit_function_time}. The fit yields the values of the six parameters $\sigma_0$, $\sigma_\infty$, $\tau_1$, $\tau_2$, $t_1^*$, and $T$ that fully define the relaxation curve. The shear stress at the transition between the two exponential steps $\sigma^*$ is computed by substituting $t=t_1^*$ into Eq.~\eqref{fit_function}. Using the resultant parameters, we find the time at which $\sigma(t)$ nearly reaches the plateau $\sigma_\infty$, denoted by $t_2^*$, by solving
\begin{equation}
    \sigma_0\exp\left(-\frac{t_2^*}{\tau(t_2^*)}\right)=\sigma_{\infty}, \label{plateau_time}
\end{equation}
where $\tau(t)$ is the same as in Eq.~\eqref{fit_function_time}.
\par

% Results
\section{Results}

\subsection{Temporal changes in stresses ($\sigma$ and $\sigma_{zz}$) and shear rate ($\dot{\gamma}$) during stress-controlled flow and flow cessation}

\begin{figure}[b]
\setlength{\abovecaptionskip}{-20pt}
\hspace*{-0.06cm}\includegraphics[scale=0.115]{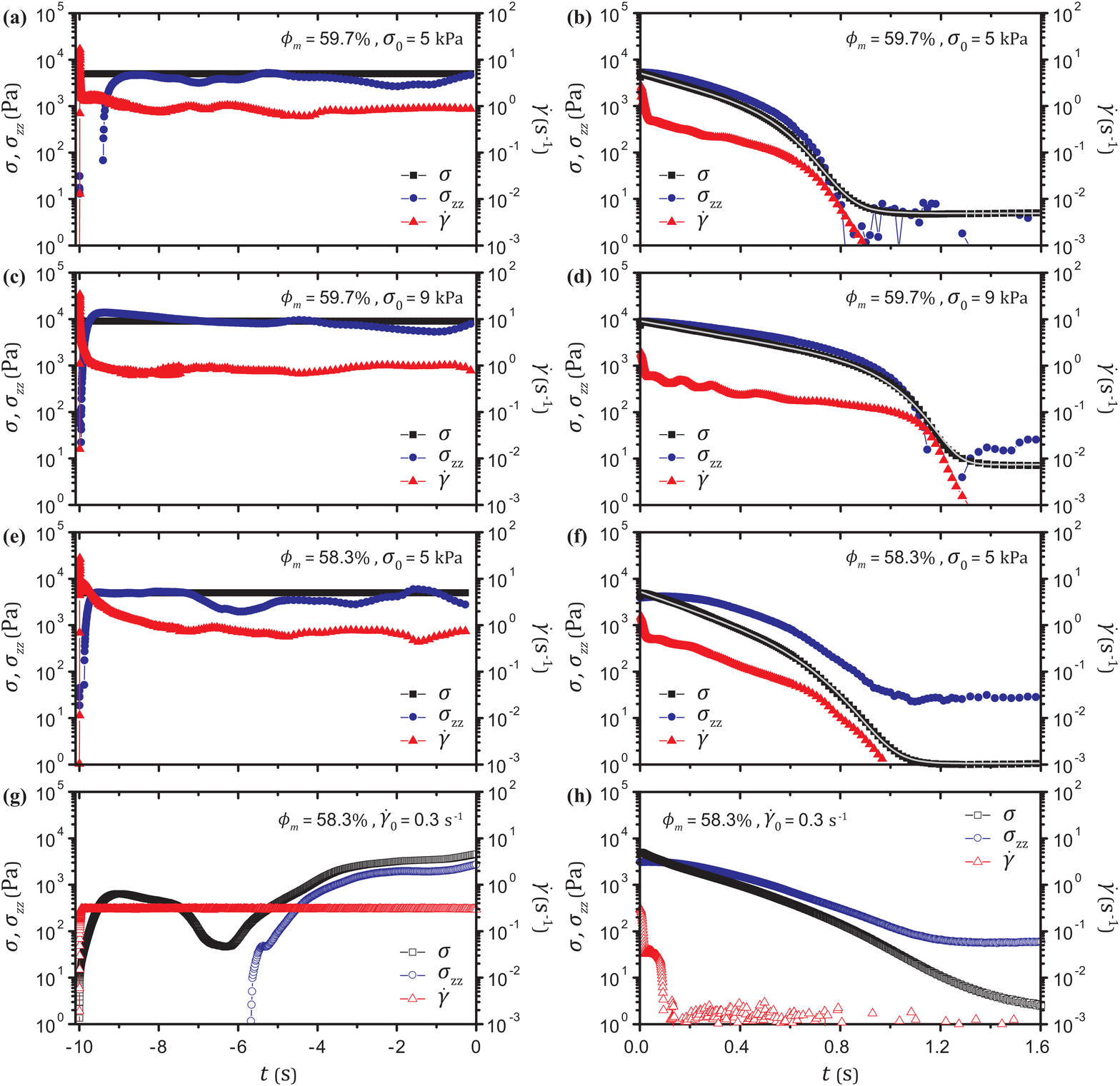}
\caption{\label{temporal} Temporal change in the shear stress $\sigma$, the normal stress $\sigma_{zz}$, and the shear rate $\dot{\gamma}$ during the stress-controlled flow (a,c,e) and upon the flow cessation (b,d,f) for samples at different effective mass fractions $\phi_m$ and different applied stresses $\sigma_0$. Gray curves in (b,d,f) are fits of Eq.~\eqref{fit_function} to the shear stress. (g) Rate-controlled flow at a shear rate $\dot{\gamma}_{0}=0.3\;\si{\per\s}$ performed on a strain-controlled rheometer, and (h) the subsequent flow cessation.}
\end{figure}

During the stress-controlled flow, the shear rate $\dot{\gamma}(t)$ exhibits an overshoot upon the application of the constant stress $\sigma_0$ at time $t=-10\;\si{\s}$, and then quickly drops to a value around which it weakly fluctuates for the remaining period of flow, as shown in Fig.~\ref{temporal}(a,c,e) for different applied stresses $\sigma_0$ and effective mass fractions $\phi_m$. The temporal mean values of $\dot{\gamma}(t)$ once $\dot{\gamma}(t)$ reaches steady state compare well with the critical shear rate $\dot{\gamma}_{c}$ observed in Fig.~\ref{flowcurves}(b). In general, the temporal mean of $\dot{\gamma}(t)$ is independent of $\sigma_0$ and decreases with the effective mass fraction $\phi_{m}$, though it shows large scatter for any given combination of $\sigma_0$ and $\phi_{m}$ due to the sensitive nature of the samples. As $\dot{\gamma}(t)$ decreases after the overshoot, the mean normal stress $\sigma_{zz}(t)$ reaches a value comparable to the applied shear stress $\sigma_0$, indicating the tendency of the system to dilate during DST \cite{Lootens2005,Brown2012,Brown2014,Royer2016,Hsiao2017,Maharjan2021}. While continuing to fluctuate, $\sigma_{zz}(t)$ also displays a temporal mean that stays constant for the rest of the flow, which indicates that the system has reached the steady state within less than $10\;\si{\s}$. The fluctuations of both $\dot{\gamma}(t)$ and $\sigma_{zz}(t)$ can be ascribed to flow instabilities such as local jamming of particles and shear banding \cite{Fall2015,Hermes2016,SaintMichel2018,Rathee2020} inherent to DST. The steady state in this work is thus loosely defined in a macroscopic sense only, and we assume that the temporal mean of $\dot{\gamma}(t)$ from $t=-5\;\si{\s}$ to $0\;\si{\s}$ represents the critical shear rate $\dot{\gamma}_c$ and equate the corresponding apparent viscosity to the steady-state viscosity $\eta_0=\sigma_0/\dot{\gamma}_c$. \par

Once the shear rate is set to zero, the shear stress $\sigma(t)$ decays exponentially in two distinct steps until it reaches a nonzero plateau at late times, as shown in Fig.~\ref{temporal}(b,d,f) along with the fits of Eqs.~\eqref{fit_function}, \eqref{fit_function_time} that satisfactorily capture $\sigma(t)$ after the flow cessation. The mean normal stress $\sigma_{zz}(t)$ closely follows $\sigma(t)$ initially and deviates from $\sigma(t)$ during the second step by decreasing to vanishingly small values that the rheometer cannot reliably measure as in Fig.~\ref{temporal}(b,d) or to a plateau as in Fig.~\ref{temporal}(f). Since a stress-controlled rheometer is used in our experiments, the solidlike stiffness of the sample in DST is reflected in a slow decrease in the shear rate $\dot{\gamma}(t)$. At the onset of the flow cessation, the shear rate $\dot{\gamma}(t)$ becomes negative and quickly rebounds to the steady-state or higher values within $0.01\;\si{\s}$, as shown in Appendix~\ref{shear_rate_rebound}. Then $\dot{\gamma}(t)$ decreases by barely an order of magnitude from the steady-state value until it rapidly diminishes at the beginning of the second step. \par

We confirm that the two-step decay in $\sigma(t)$ is an inherent behavior of the DST fluids, rather than an artifact stemming from the slow decrease in $\dot{\gamma}(t)$ by detecting the two-step relaxation on a strain-controlled rheometer (ARES-G2, TA Instruments) that is capable of keeping $\dot{\gamma}(t)$ below $0.01\;\si{\per\s}$ after the first $0.1\;\si{\s}$, as shown in Fig.~\ref{temporal}(h). Unlike the data collected from the stress-controlled rheometer, the one from the strain-controlled rheometer displays a momentary decrease in $\sigma(t)$ at the very beginning, since the abrupt reduction in $\dot{\gamma}(t)$ leads to a corresponding reduction in the hydrodynamic stress \cite{Morris2020}. The two-step exponential relaxation, however, ensues this initial stress drop, corroborating the validity of the data from the stress-controlled rheometer. Despite its effective rate control, the strain-controlled rheometer does not allow the rapid switch from applying a stress boundary condition (constant stress) during the initial flow to applying a kinematic boundary condition (zero shear rate) during the flow cessation. Although the two-step relaxation is accessible instead by initially applying a constant shear rate, rather than a constant stress, the variability of the critical shear rate of any given sample and the large range of stresses that correspond to the single critical shear rate prevent us from exploring the states that exhibit different stresses at DST by imposing a shear rate. This lack of control of the shear-thickened state by a strain-controlled protocol is evident in the larger and continuous fluctuations of $\sigma(t)$ and $\sigma_{zz}(t)$, as shown in Fig.~\ref{temporal}(g). We therefore use the results from the stress-controlled rheometer only for the following analysis. \par

\subsection{Dependence of relaxation parameters on the steady-state viscosity $\eta_0$}

\begin{figure}[t]
\setlength{\abovecaptionskip}{-20pt}
\hspace*{-0.06cm}\includegraphics[scale=0.12]{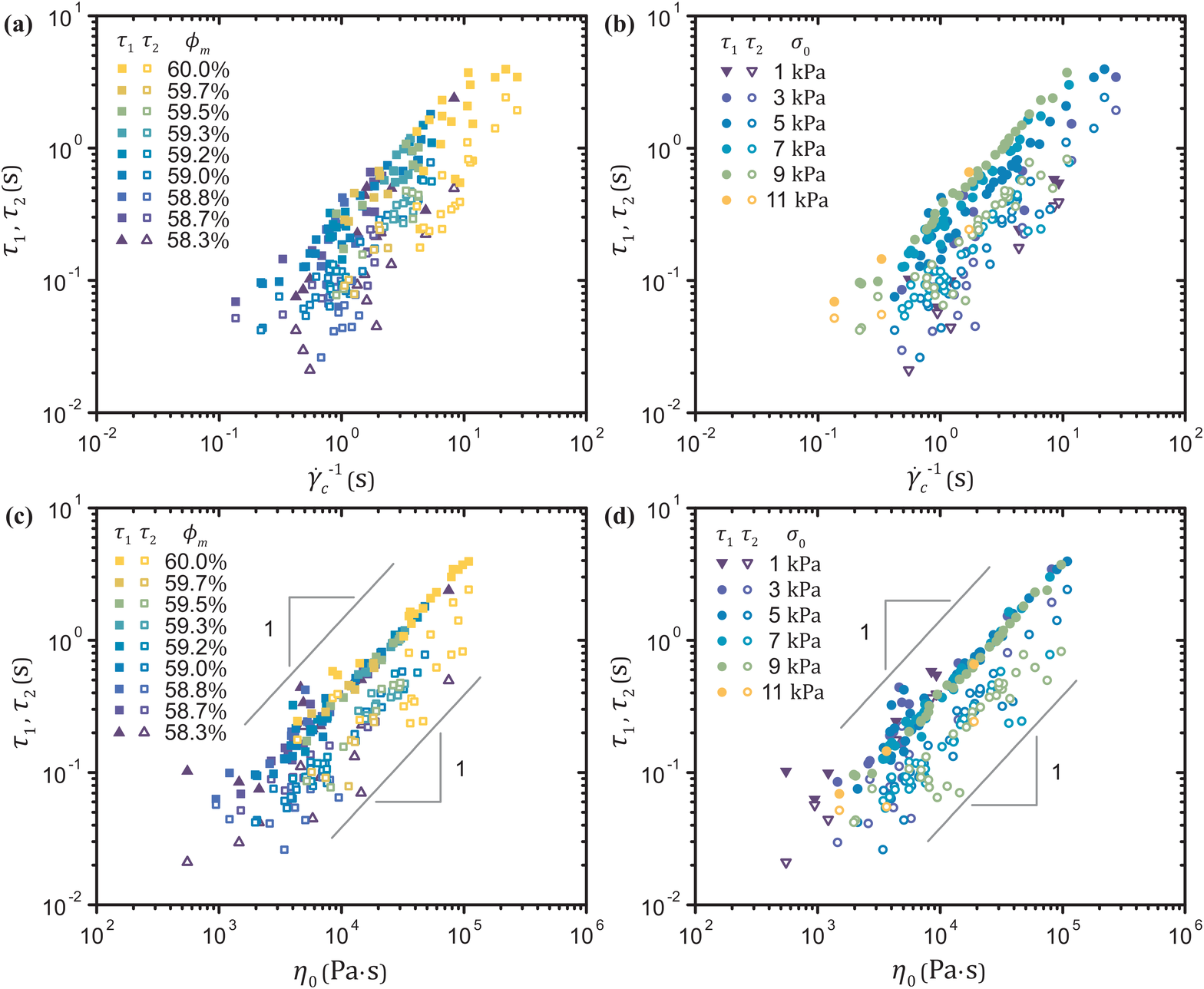}
\caption{\label{tau1tau2} Characteristic times $\tau_1$ (filled symbols) and $\tau_2$ (open symbols) as a function of the reciprocal of the critical shear rate $\dot{\gamma}_c^{-1}$ and the steady-state viscosity $\eta_0$. Different colors represent different effective mass fractions $\phi_m$ in (a,c), and different applied stresses $\sigma_0$ in (b,d). The data points that correspond to the lowest $\phi_m$ and the lowest $\sigma_0$, denoted by triangles and inverted triangles respectively, deviate from the linear relationship between either characteristic time and $\eta_0$ displayed by all the other data sets.}
\end{figure}

The characteristic times of both the first and the second exponential decays of $\sigma(t)$, $\tau_1$ and $\tau_2$ respectively in Eq.~\eqref{fit_function_time}, increase with the reciprocal of the critical shear rate ${\dot{\gamma}_c}^{-1}$, as shown in Fig.~\ref{tau1tau2}(a). The reciprocal critical shear rate ${\dot{\gamma}_c}^{-1}$ can serve as a measure of the actual mass fraction of each sample as discussed in Ref.~\cite{Maharjan2017}. We confirm that higher effective mass fractions $\phi_m$ correspond to higher ${\dot{\gamma}_c}^{-1}$ although large scatters are present in Fig.~\ref{tau1tau2}(a). Labeling the data points with different colors for different applied stresses $\sigma_0$ as in Fig.~\ref{tau1tau2}(b), however, reveals a systematic dependence of $\tau_1$ and $\tau_2$ on $\sigma_0$, an increase of which shifts both characteristic times to lower ${\dot{\gamma}_c}^{-1}$. This $\sigma_0$ dependence is factored out when $\tau_1$ and $\tau_2$ are plotted as a function of the steady-state viscosity $\eta_0=\sigma_{0}{\dot{\gamma}_c}^{-1}$ as in Fig.~\ref{tau1tau2}(c,d), which elucidates that $\tau_1$ and $\tau_2$ scale linearly with $\eta_0$. Since $\eta_0$ tends to increase with both the effective mass fraction $\phi_m$ and the applied stress $\sigma_0$, we can interpret $\eta_0$ as a quantity that accounts for the effects of both $\phi_m$ and $\sigma_0$. The data points from the experiments at the lowest $\phi_{m}=58.3\%$ and $\sigma_0=1\;\si{\kilo\Pa}$ exhibit significant deviation in Fig.~\ref{tau1tau2}(c,d), which illustrates that the dependence of $\tau_1$ and $\tau_2$ on $\eta_0$ marks a behavior of the dense suspensions well into the DST regime only. Maharjan and Brown report that multistep relaxation is observed at higher mass fractions than the one at which DST starts to occur \cite{Maharjan2017}, and our results suggest that the $\eta_0$ dependence of the characteristic times requires the system to be even deeper within the DST regime. To focus on the behavior of the samples deeper into DST, the data sets corresponding to $\phi_{m}=58.3\%$ and $\sigma_0=1\;\si{\kilo\Pa}$ are omitted in the following analysis. The greater amount of noise in $\tau_2$ than in $\tau_1$ can be partially attributed to the lower magnitudes of $\tau_2$, but mainly results from the overall shorter duration of the second relaxation, which renders the late-time plateau of $\tau(t)$ in Eq.~\eqref{fit_function_time} rather elusive. \par

\begin{figure}[t]
\setlength{\abovecaptionskip}{-20pt}
\hspace*{-0.06cm}\includegraphics[scale=0.12]{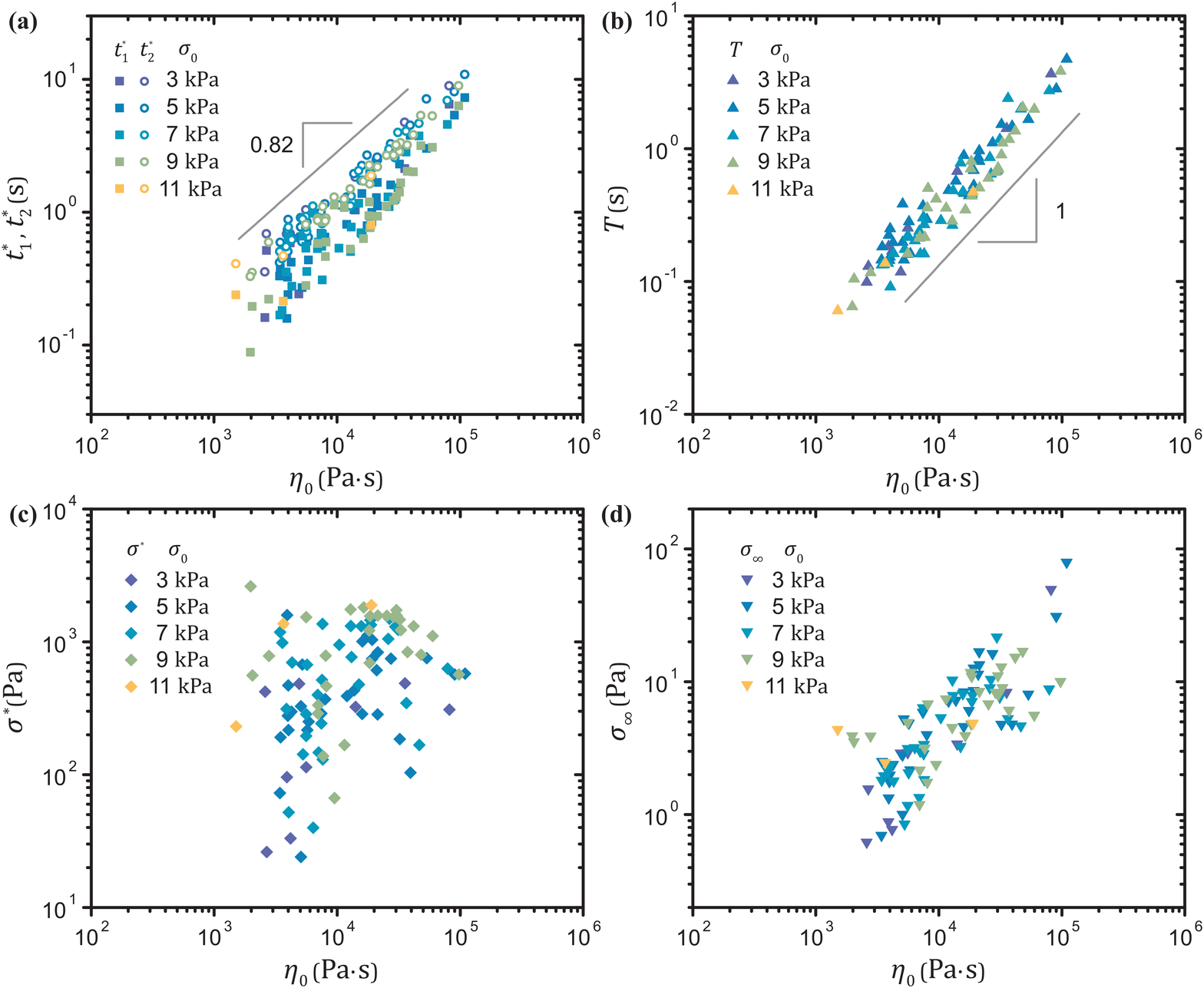}
\caption{\label{t1t2} (a) Transition times into the second exponential relaxation $t_1^*$ (filled squares) and into the late-time plateau $t_2^*$ (open circles) as a function of the steady-state viscosity $\eta_0$. (b) Scale parameter $T$ of the complementary error function, (c) stress $\sigma^*$ at the transition into the second exponential step, and (d) residual shear stress $\sigma_{\infty}$ as a function of $\eta_0$. Different colors represent different applied stresses $\sigma_0$.}
\end{figure}

All the other time parameters of our fit also exhibit nearly linear dependence on the steady-state viscosity $\eta_0$, as displayed in Fig.~\ref{t1t2}(a,b). The transition time $t_1^*(\eta_0)$ into the second exponential decay of $\sigma(t)$ obeys a power law with an exponent $0.82\pm0.04$, while the transition time  $t_2^*(\eta_0)$ into the late-time plateau $\sigma_{\infty}$ exhibits a power-law exponent of $0.82\pm0.02$. Even the scale parameter $T$ of the complementary error function in Eq.~\eqref{fit_function_time} increases linearly with $\eta_0$. By contrast, the value of the shear stress at the transition from the first to the second step $\sigma^*=\sigma(t=t_1^*)$ displays no clear dependence on the steady-state viscosity $\eta_0$, as shown in Fig.~\ref{t1t2}(c). The transition normal stress $\sigma_{zz}^*=\sigma_{zz}(t=t_1^*)$ follows nearly identical trends as those found for $\sigma^*$, as presented in Appendix~\ref{transition_normal_stress}, consistent with the observation that $\sigma(t)$ and $\sigma_{zz}(t)$ exhibit similar behaviors until the beginning of the second step. We note that $\sigma^*$ far exceeds the stress corresponding to the onset of DST in the steady-state experiments, and thus that the first step of the relaxation ends while the system is still deep in the DST regime. Remarkably, also the residual shear stress $\sigma_{\infty}$ at the end of the relaxation, albeit noisy, increases with the steady-state viscosity $\eta_0$ nearly linearly. Since $\eta_0$ represents the fraction of frictional contacts generated during the initial flow, this positive correlation between $\sigma_{\infty}$ and $\eta_0$ indicates that even the second step of the relaxation does not erase the memory of DST. \par

% Discussion
\section{Discussion}
The robust dependence of all the fitting parameters of the two-step stress relaxation on the steady-state viscosity $\eta_0$, which is directly linked to the extent of the frictional interparticle interactions, hints at continued existence of frictional force chains generated before the flow cessation. An interpretation of the critical shear rate $\dot{\gamma}_c$ to scale as the reciprocal of an ``intrinsic contact relaxation time'' \cite{Melrose2004} between particles prompts a hypothesis that one of the characteristic timescales represents the time over which particles lose frictional contacts \cite{Maharjan2017}. Yet the evident dependence of the characteristic times on $\eta_0=\sigma_0\dot{\gamma}_c^{-1}$ rather than on $\dot{\gamma}_c^{-1}$, clarifies that it is not such an intrinsic contact relaxation time that governs the stress relaxation upon flow cessation. The relaxation hence does not necessarily manifest relubrication of the particles that comprise the force chains. It, nonetheless, does involve continuous rearrangements of the internal structure of the material at the particle level, as evidenced by the diminishing structural anisotropy that has been optically measured in DST suspensions composed of polymethylmethacrylate (PMMA) particles \cite{DHaene1993}. These rearrangements inevitably vary how the contact network of particles formed in DST bears the external stress. The presence of the two distinct relaxation steps thus leads us to hypothesize that the two steps are governed by two different types of particle motions that do not recover all interparticle lubrication layers. \par

An obvious feature that distinguishes the first step of the relaxation from the second one is that the suspension exhibits pronounced resistance against the stress relaxation, as evidenced by the high values of the mean normal stress $\sigma_{zz}(t)$. The normal stress $\sigma_{zz}(t)$ is a measure of the tendency of the system to dilate, and thus that of the frustration of the particle network due to the fixed gap size between the two plates. That $\sigma_{zz}(t)$ remains significantly high during the first relaxation dictates that the microstructure remains largely frustrated and its macroscopic elasticity persists. A simple analogy to the linear viscoelastic Maxwell model enables us to estimate the effective elastic modulus $G$ of the system, as a Maxwell fluid exhibits an exponential stress decay during flow cessation $\sigma(t)=\sigma_0\exp(-Gt/\eta)$, where $\eta$ denotes that characteristic viscosity. The elastic modulus $G=\eta_0/\tau_1\approx27\;\si{\kilo\pascal}$ estimated based on our results is much lower than the modulus of cornstarch particles of the order of $1\;\si{\giga\pascal}$ \cite{Han2017}, but higher than the applied stresses $\sigma_0=3-11\;\si{\kilo\pascal}$, which attests to the prolonged stiffness of the system. Although such direct comparison between the ideal Maxwell fluid and the DST fluids needs to be made with caution because of the nonlinearity and the plasticity present in the system response, we note that the rate of relaxation in the first step captured in a more sophisticated model of Ref.~\cite{Baumgarten2020} is also set by the elastic modulus. We thus infer that, during the first step of the relaxation, the particles move collectively to induce elastic structural relaxation of the contact network, but its plastic disintegration is minimal. \par

The rapid reduction in the shear stress $\sigma(t)$, the normal stress $\sigma_{zz}(t)$, and the shear rate $\dot{\gamma}(t)$ initiated as the relaxation transitions into the second step indicates that significant particle rearrangements occur within the network. The decoupling between the mean normal stress $\sigma_{zz}(t)$ and the shear stress $\sigma(t)$ marks a fading feature of DST \cite{Royer2016,Maharjan2021}, suggesting incipient breakage of the contact network due to interparticle repulsions. Although considerably scattered, the transition stress $\sigma^*$ is largely independent of $\eta_0$ for all of our experiments, as shown in Fig.~\ref{t1t2}(c). Since $\eta_0$ contains the information about both the effective mass fraction $\phi_m$ and the applied stress $\sigma_0$, the $\eta_0$ independence of the transition stress $\sigma^*$ may mirror the effects of the interparticle repulsions that depend on neither $\phi_m$ nor $\sigma_0$. The scatter in the data, which prevents a precise measurement of $\sigma^*$, may result from that the mean normal stress $\sigma_{zz}(t)$, and equivalently the shear stress $\sigma(t)$, serves as an indirect measure only of the particle pressure $p=-\frac{1}{3}\tr(\boldsymbol{\sigma})$, where $\boldsymbol{\sigma}$ denotes the stress tensor, the determination of which necessitates normal stress measurements also in the radial and the azimuthal directions. Furthermore, the inherent spatiotemporal heterogeneity of the stress carried by the contact network \cite{Rathee2020,SaintMichel2018,Hermes2016}, as well as the nonuniform strain imposed in the plate-plate geometry, limits the resolution of the information we can obtain from the averaged stress measurements in conventional rheometry. Nevertheless, the description of the second relaxation step as the onset of network breakage is in line with the prediction of the constitutive model in Ref.~\cite{Baumgarten2020}, where the second relaxation time is set by one of the plasticity constants. \par

The dependence of the residual stress $\sigma_\infty$ on the steady-state viscosity $\eta_0$, however, shows that the system retains its memory of DST even after the second relaxation step. Given the sensitivity of $\eta_0$ to the amount of frictional contact \cite{Wyart2014} prior to the flow cessation, the lasting effect of $\eta_0$ on $\sigma_\infty$ demands that the cornstarch suspension does not fully revert to the state before DST. This result is once again consistent with the prediction of the model in Ref.~\cite{Baumgarten2020}, which relates the two-step relaxation to the fraction of frictional contacts staying at a nonzero value even at the end of the relaxation. The difference in the pathways taken by the system into and out of DST may be a manifestation of the hysteresis of rate-controlled flows reflected in the S-shaped stress-rate ($\sigma-\dot{\gamma}$) curve that yields a lower critical shear rate for a frictional-to-lubricated transition than for a lubricated-to-frictional transition \cite{Brown2012,Wyart2014,Pan2015}. The prominence of the hysteretic response of dense cornstarch suspensions is indeed known to give rise to peculiar dynamical behaviors, such as the formation of persistent holes during vibration \cite{Merkt2004,Deegan2010} and the nonmonotonic settling of a solid sphere through a bath of the suspension \cite{vonKann2011}. Yet, in flow cessation of the dense suspensions, such hysteresis cannot fully account for the lasting effect of DST until the very end of the flow, because the frictional-to-lubricated transition has to occur at a finite shear rate before it drops to zero. Instead, we mainly attribute the persistence of frictional contact to the stationary boundaries in the absence of flow, imposed by the zero shear rate $\dot{\gamma}(t)=0$. Since the structure of the contact network has evolved to adapt to the relative motion of the boundaries by interlocking the particles with a corresponding directionality \cite{Cates1998}, it takes the opposite motion of the boundaries for the system to fully relax, as reported in studies where flow reversal brings the stress back to zero after shear thickening \cite{GadalaMaria1980,Lin2015}. Stopping the upper plate of the rheometer in the flow cessation thus keeps the contact network partially frustrated, causing the retention of the memory of DST. Although the interparticle repulsive forces can induce partial breakdown of the force chains, which we infer to be the main cause of the rapid decrease in the stresses during the second relaxation, the limited deformability of the particles and the presence of adhesion between particles triggered by frictional contact \cite{Galvez2017,James2018,Hsu2021} hinder relubrication of the particles and consequently a complete relaxation. This inability of the DST fluid to relax contrasts the behavior observed in a suspension of jammed soft, Brownian particles upon flow cessation \cite{Mohan2013}, where the considerable deformability of the particles, as well as thermal fluctuations, facilitates collective motions of the particles, leading to the slow, but continuous decrease in the stress. \par

% Conclusion
\section{Conclusions}
We show that the time parameters that characterize the two-step exponential stress relaxation in dense aqueous cornstarch suspensions upon flow cessation after discontinuous shear thickening (DST) are determined by the steady-state apparent viscosity $\eta_0$ during DST. Additionally, we demonstrate that the residual stress $\sigma_\infty$ at the end of the flow cessation is also positively correlated to $\eta_0$, which is a proxy for the fraction of frictional contacts of all particle interactions. This dependence of the relaxation parameters of the shear stress $\sigma(t)$ on $\eta_0$ is robust as long as the system is initially sheared well into the DST regime at sufficiently high applied stresses $\sigma_0$ and effective mass fractions $\phi_m$. The memory of DST hence lingers even after the flow stops, which suggests that the particles remain in frictional contact despite the repulsive forces between them. Based on the high normal stress $\sigma_{zz}(t)$, we infer that the frictional force chains during the first step undergo primarily viscoelastic relaxation without considerable disintegration of the network structure. During the second step, the ensuing steep decrease in the shear stress $\sigma(t)$ accompanied by its decoupling from the mean normal stress $\sigma_{zz}(t)$ marks the onset of the plastic breakage, due to the interparticle repulsions, of the microstructure that can sustain high internal stresses. The lasting effect of the steady-state viscosity $\eta_0$ on the nonzero final stress $\sigma_\infty$, however, corroborates an incomplete frictional-to-lubricated transition, which may stem from the persistent frustration of the contact network under fixed kinematic boundary conditions and the adhesion between cornstarch particles activated upon frictional contact. \par

\begin{acknowledgments}
We acknowledge support from the MIT Research Support Committee and Kwanjeong Educational Foundation, Awards No. 16AmB02M and No. 18AmB59D (J.H.C.), the U.S. Air Force ROTC STEM Graduate Scholar Program (A.H.G.), the Royal Society, Exchange Grant No. IES{\textbackslash}R2{\textbackslash}170104 (I.R.P. and I.B.), and MISTI Global Seed Funds Award (I.B. and I.R.P.).
\end{acknowledgments}

%Appendices
\appendix
\section{Initial response of the shear rate $\dot{\gamma}(t)$ upon flow cessation}
\label{shear_rate_rebound}
\begin{figure}[h]
\setlength{\abovecaptionskip}{-20pt}
\hspace*{-0.06cm}\includegraphics[scale=0.12]{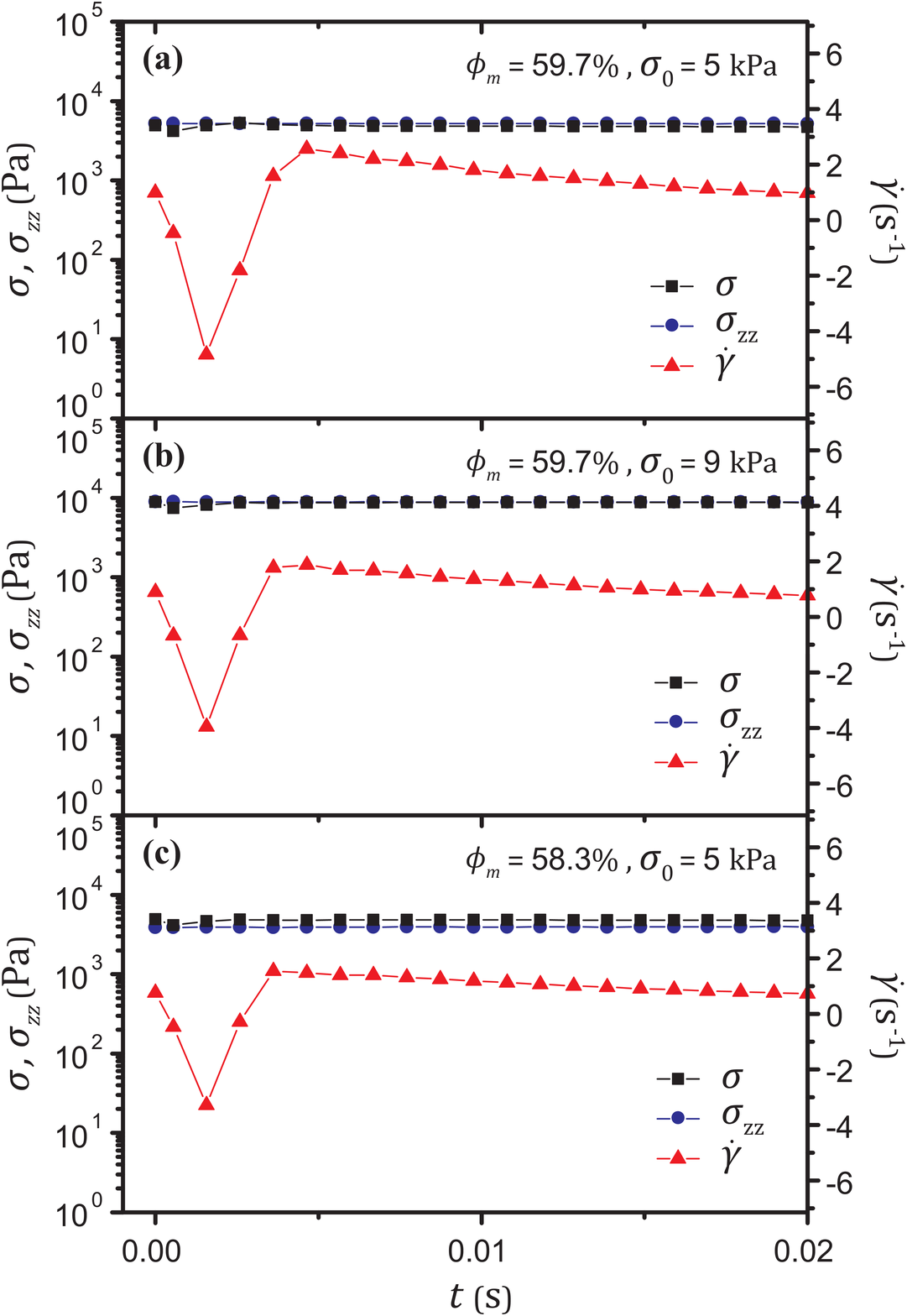}
\caption{\label{gamma_dot_rebound} Temporal change in the shear stress $\sigma$, the normal stress $\sigma_{zz}$, and the shear rate $\dot{\gamma}$ for the first $0.02\;\si{\s}$ upon flow cessation for samples at different effective mass fractions $\phi_m$ and different applied stresses $\sigma_0$. The data are identical to the ones in Fig.~\ref{temporal}, but the shear rate $\dot{\gamma}$ is plotted in linear scale to show the negative values of $\dot{\gamma}$ as the rheometer attempts to stop the flow at $t=0\,\si{\s}$.}
\end{figure}

At the onset of flow cessation, the shear-thickening suspensions exhibit pronounced elastic responses where the shear rate $\dot{\gamma}(t)$ drops to negative values and rebounds to the steady-state or higher values within $0.01\;\si{\s}$, as displayed in Fig.~\ref{gamma_dot_rebound}.

\section{Transition normal stress $\sigma_{zz}^*$}
\label{transition_normal_stress}
\begin{figure}[h]
\setlength{\abovecaptionskip}{-20pt}
\hspace*{-0.06cm}\includegraphics[scale=0.12]{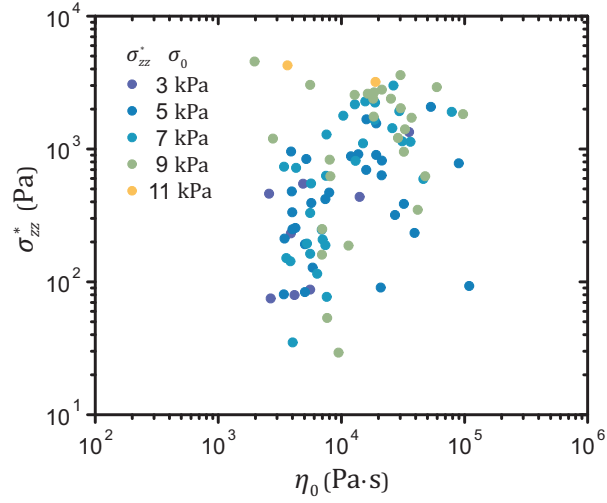}
\caption{\label{normal_stress_star} Transition mean normal stress $\sigma_{zz}^*=\sigma_{zz}(t=t_1^*)$ as a function of the steady-state apparent viscosity $\eta_0$.}
\end{figure}

Similar to the transition shear stress $\sigma^*$, the mean normal stress at the transition $\sigma_{zz}^*=\sigma_{zz}(t=t_1^*)$ from the first to the second step of the exponential decay of the shear stress $\sigma(t)$ exhibits no systematic dependence on the steady-state viscosity $\eta_0$, as displayed in Fig.~\ref{normal_stress_star}. The transition normal stress $\sigma_{zz}^*$ also shows significant scatter.

\newpage
\bibliography{Dense_suspension_relax_Aug_21_fixed.bib}

\end{document}